\documentclass{aa}  

\usepackage{graphicx}
\usepackage{txfonts}
\begin{document} 

    \title{Point source detection with fully convolutional networks}
    \subtitle{Performance in realistic microwave sky simulations}
    \titlerunning{PoSeIDoN in Realistic Simulations}
    \authorrunning{Bonavera L. et al.}
   
    \author{Bonavera L.\inst{1,2},
          Suarez Gomez S. L.\inst{2,3},
          Gonz{\'a}lez-Nuevo J.\inst{1,2},
          Cueli M. M.\inst{1,2},
          Santos J.D.\inst{1,2}, Sanchez M.L.\inst{1,2},\\Muñiz R. \inst{4}, de Cos F.J.\inst{2,5}}

   \institute{$^1$Departamento de F{\'i}sica, Universidad de Oviedo, C. Federico Garc{\'i}a Lorca 18, 33007 Oviedo, Spain\\
             $^2$Instituto Universitario de Ciencias y Tecnolog{\'i}as Espaciales de Asturias (ICTEA), C. Independencia 13, 33004 Oviedo, Spain\\
             $^3$Departamento de Matematicas, Universidad de Oviedo, C. Federico Garcia Lorca 18, 33007 Oviedo, Spain\\
             $^4$Departamento de Inform{\'a}tica, Universidad de Oviedo, Edificio Departamental 1. Campus de Viesques s/n, 33204 Gij{\'o}n, Spain\\
             $^5$Escuela de Ingeniería de Minas, Energ{\'i}a y Materiales, Independencia 13, 33004 Oviedo, Spain}


  \abstract
   {Point sources are one of the main contaminants to the recovery of the cosmic microwave background signal at small scales, and their careful detection will be important for the next generation of cosmic microwave background experiments like LiteBird.}
   {We want to develop a method based on fully convolutional networks to detect sources in realistic simulations, and to compare its performance against one of the most used point source detection method in this context, the Mexican hat wavelet 2 (MHW2). The frequencies for our analysis are the 143, 217, and 353 GHz Planck channels.}
   {We produce realistic simulations of point sources at each frequency taking into account potential contaminating signals as the cosmic microwave background, the cosmic infrared background, the Galactic thermal emission, the thermal Sunyaev-Zel'dovich effect, and the instrumental and point source shot noises. We first produce a set of training simulations at 217 GHz to train the neural network that we named PoSeIDoN. Then we apply both PoSeIDoN and the MHW2 to recover the point sources in the validating simulations at all the frequencies, comparing the results by estimating the reliability, completeness, and flux density estimation accuracy. Moreover, the receiver operating characteristic (ROC) curves are computed in order to asses the methods'performance.}
   {In the extra-galactic region with a 30$^\circ$ galactic cut, the neural network successfully recovers point sources at 90\%  completeness corresponding to 253, 126, and 250 mJy for 143, 217, and 353 GHz respectively. In the same validation simulations the wavelet with a 3$\sigma$ flux density detection limit recovers point sources up to 181, 102, and 153 mJy at   90\% completeness. To reduce the number of spurious sources, we also apply a safer 4$\sigma$ flux density detection limit, the same as in the \textit{Planck} catalogues, increasing the 90\%   completeness levels: 235, 137, and 192 mJy. In all cases PoSeIDoN produces a much lower number of spurious sources with respect to MHW2. As expected, the results on spurious sources for both techniques worsen when reducing the galactic cut to 10$^\circ$.}
   {Our results suggest that  using  neural networks is a very promising approach for detecting point sources using data from cosmic microwave background experiments, providing overall better results in dealing with spurious sources with respect to the more usual filtering approaches. Moreover, PoSeIDoN gives competitive results even at the 217 GHz nearby channels where the network was not trained.}


   \maketitle
%

\section{Introduction}

The importance of compact sources (galaxy clusters and extra-galactic sources) for ground- and space-based cosmic microwave background (CMB) experiments has been clear since the conception of the  Wilkinson Microwave Anisotropy Probe (WMAP) \citep{BEN13} and Planck \citep{PLA18_I} missions. Point sources (PSs) in the microwave regime are mainly blazars (i.e. AGNs with the relativistic jets aligned along the line of sight) and dusty galaxies. 
At such frequencies, PSs are one of the contaminants to the recovery of the CMB anisotropy signal whose effect is more important at small angular scales. For this reason PSs are even more important for the next generation of CMB experiments with higher resolution than \textit{Planck}, such as the Cosmic Origins Explorer \citep[CORE,][]{Del18}, the Probe of Inflation and Cosmic Origins \citep[PICO,][]{Han19}, or the Lite satellite for the studies of B-mode polarisation and Inflation from cosmic background Radiation Detection \citep[LiteBIRD,][]{Mat14}. Generally, they are planned to keep the PS contamination low (e.g. choosing a frequency range around the minimum PS contribution, selecting sky areas with deep extragalactic surveys available, and/or planning for simultaneous PS observations to minimise the impact of variable sources), but precise CMB measurements will still be affected by PSs. For this reason, it is  important to develop highly performing methods for PS detection. 

The standard single-frequency approach for PS detection in the CMB and far-IR frequencies rely on the Mexican hat wavelet \citep[MHW; ][]{VIE03,GN06} or on the matched filtering techniques \citep{TEG98,BAR03,LOP06, Her02}. A matched filter is theoretically the optimal filter when the PS shape is known providing the maximum signal-to-noise amplification. However, as concluded by \citet{LOP06}, the second member of the MHW family \citep[MHW2; ][]{GN06} provides a similar performance to  matched filtering, but it is easier to implement and more robust. This wavelet has been successfully applied to \textit{Planck} realistic simulations \citep{GN06,LOP06,LEA08} and    to WMAP \citep{LOP07,MAS09} and \textit{Planck} real data: the Early Release Compact Source Catalogue \citep[ERCSC, ][]{ERCSC}, the \textit{Planck} Catalogue of Compact Sources \citep[PCCS, ][]{PCCS}, and the Second \textit{Planck} Catalogue of Compact Source \citep[PCCS2, ][]{PCCS2}. This is why we decided to compare our results against this method.

Although there was always a tight relationships between machine learning techniques and astrophysics and cosmology, in  recent years the particular usage of neural networks has become a mainstream technique to derive new results. Artificial neural networks are artificial intelligence (AI) techniques involving numerical mathematical models which can be trained to represent complex physical systems by supervised or unsupervised learning \citep{SUA19a, SUA17}. These characteristics are perfect to provide further results in cosmology. Some example of recent interesting applications of artificial neural networks in cosmology are the identification of galaxy mergers \citep{Pea19} and strongly gravitational lenses \citep{Pet17} in astronomical images, a better estimation of cosmological constraints from weak lensing maps \citep{Flu19} and high fidelity generation of weak lensing convergence maps \citep{Mus19}, and cosmological structure formation simulations under different assumptions \citep{Mat18, He19, Per19, Gui19}.

Some AI approaches, such as the multi-layer perceptron (MLP) \citep{JUE12} or convolutional neural networks (CNNs) \citep{SUA19b, KRI12}, have been successfully applied to image processing (and related fields) for modelling and forecasting \citep{GRA13, GIU13}. In this work we propose the use of fully convolutional networks (FCNs) \citep{LON15, DAI16} as a very promising tool for PS detection. They are usually applied in image recognition, and make use of different layers in order to get various image features (e.g. shapes, smoothness, and borders). Important features of the input images are generally obtained by pairing convolution and merging layers. After that, layers are applied to get the output (image or numerical). In this work we present an application of FCNs to the detection of PSs in realistic simulations, the Point Source Image Detection Network (\mbox{PoSeIDoN}) that can be summarised in the search of compact sources in a noisy background (i.e. the rest of the components in the microwave sky).

The outline of the paper is the following. Section \ref{sec:simulations} describes how the simulated maps are generated and Section \ref{sec:methodology} reviews our methodology. The results are presented in Section \ref{sec:results}, and our conclusions are in Section \ref{sec:conclusions}.


\section{Simulations}
\label{sec:simulations}

   \begin{figure*}[ht]
   \centering
   \includegraphics[width=19cm]{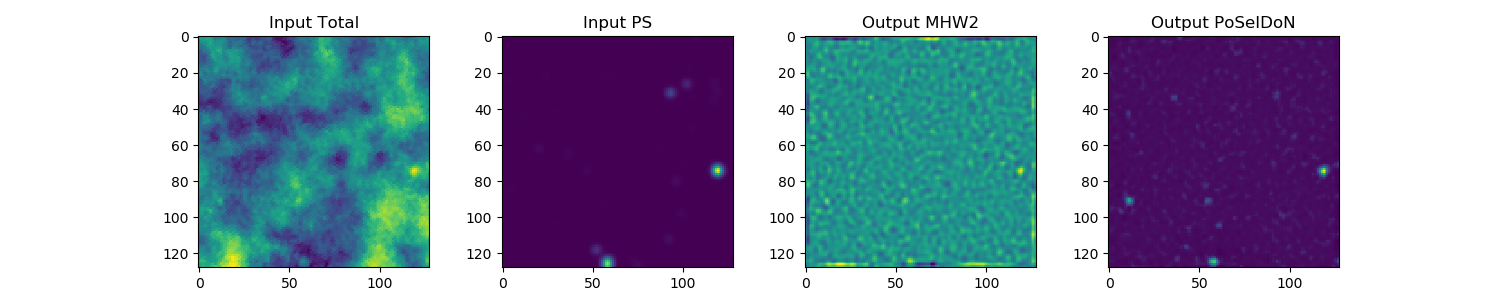}
   \includegraphics[width=19cm]{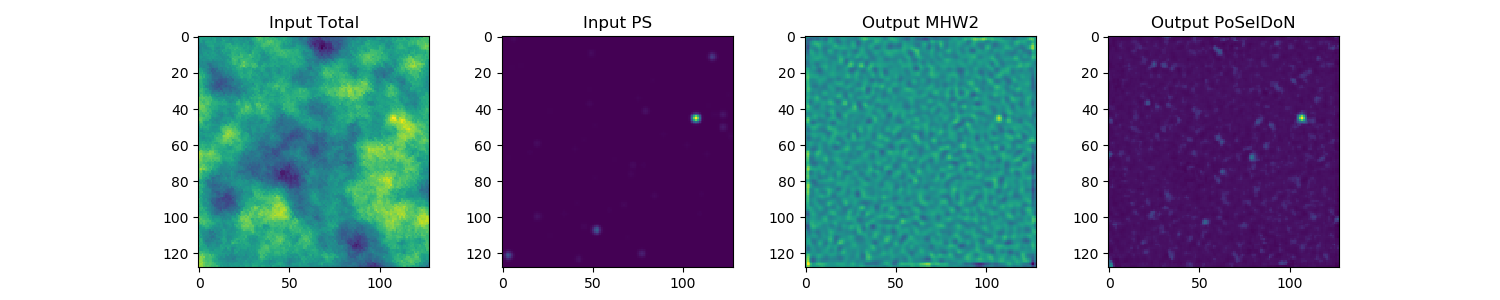}
   \includegraphics[width=19cm]{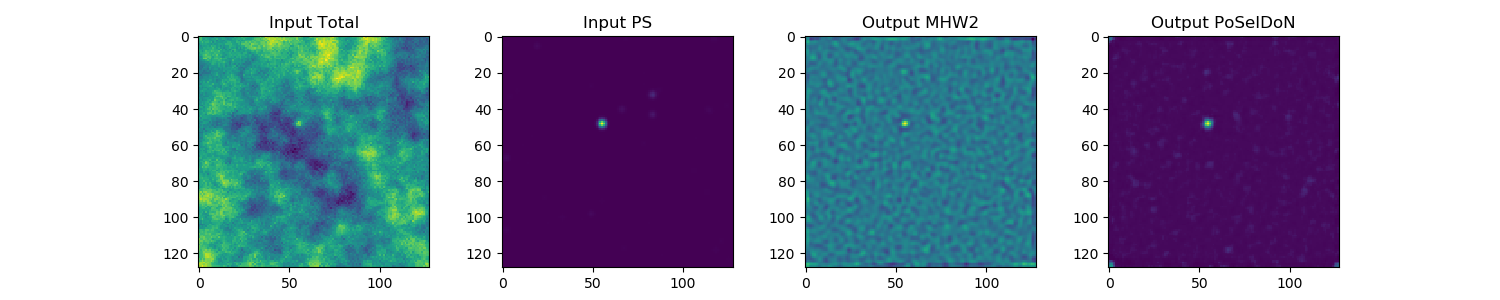}
      \caption{From left to right: Sample patch comparison among the total and PS input validation maps and the MHW2 and PoSeIDoN PS outputs, for 143, 217, and 353 GHz (top, middle, and bottom panel, respectively) at $b>30^\circ$ Galactic latitudes. The number, position, and flux density of the PSs are different at each frequency.}
         \label{Fig:patches}
   \end{figure*}
In this work we make use of realistic simulated maps of the microwave sky. The simulations correspond to sky patches at 143, 217, and 353 GHz, the central channels of the \textit{Planck} mission, with $pixsize=90$ arcsec, a round number close to the 1.72 arcmin used in the \textit{Planck} maps \citep[npix=2048 in the \texttt{HEALPIX} all-sky pixelisation schema;][]{GOR05}. For reasons of memory and speed, we use patches of $128\times128$ pixels, a trade-off between density of bright sources per patch and size. We also tested   that using bigger patches ($256\times256$) does not alter our statistical results or our conclusions.

First, a catalogue of radio PSs is simulated at each frequency independently by following the model by \citet{TUC11}. The flux density limit is $1$ mJy at all the frequencies. From the simulated catalogue we then create the simulated PS map and convolve it with the FWHM of the instrument \citep[7.22, 4.90, and 4.92 arcmin at 143, 217, and 353 GHz, respectively;][]{PLA18_I}. 

In order for the simulations to be realistic at these frequencies, we need to take into account fluctuations due to high redshift infrared PSs \citep[massive proto-spheroidal galaxies in the process of forming most of their stellar mass;][]{Gra04, Lap06, LAP11, CAI13} too faint to be detected one by one. Such contamination \citep{Bla98, Lag03, Dol04} is dominant at resolutions of a few arcmin , and   is called the cosmic infrared background \citep[CIB;][]{Pug96,Hau01,Dol06}. We use the software \texttt{CORRSKY} \citep{GN05} to simulate a sample of galaxies with a particular clustering properties, described by their angular power spectrum \textit{P(k)}. We adopt the power spectrum and the source number counts (different at each frequency) given by the \cite{LAP11} and \cite{CAI13} models, respectively.

Radio and late-type infrared galaxies \citep{Tof98,PEPXIII, PIPVII, PCCS} are generally only distinguishable for their different spectral emission, not for their shape and/or size. Compared with the \textit{Planck} beam they are in general both point-like sources. Their introduction supposes simply a higher density of brighter PSs {per} patch in the highest simulated frequency without appreciably modifying the statistical properties of the background. However, in order to also take into account  their shot-noise contribution, we have simulated them (randomly distributed in the sky) using the source number counts \cite{CAI13} model normalised to the later updated \citet{Neg13} bright source density observed levels. For all the source populations (radio, late-type, and CIB) and frequencies, we simulate them down to the same flux density limit of 3 $\mu$Jy.

On larger angular scales we must include in our simulation the contamination due to diffuse emission by our Galaxy and the CMB. These contaminants are introduced in our simulated maps by randomly select patches in \textit{Planck} 143, 217, and 353 GHz official CMB maps \citep[from the last release described in ][]{PLA18_I}. The CMB maps are the ones by the SEVEM method \citep{LEA08, FER12}, and are provided at all \textit{Planck} frequencies. For the Galaxy emission we use the \textit{Planck} Legacy Archive (PLA) simulations\footnote{available at http://pla.esac.esa.int/pla/}, available for all \textit{Planck} channels. These simulations were produced using the Planck Sky Model software \citep{Del13}. The \textit{Planck} maps are at nside=2048, which corresponds to a pixel size of 1.72 arcmin, and the selected sky patches\footnote{The central position of each selected sky patch is randomly chosen, and the probability of repeating exactly the same sky position is negligible.}
are projected into flat patches with pixel size of 1.5 arcmin using the \texttt{gnomview} function provided with the \texttt{HEALPIX} framework \citep{GOR05}. We note that,  as the original templates used to simulate the thermal dust emission had worse resolution than the \textit{Planck} template at these frequencies, small-scale fluctuations were added following the \citet{Miv07} method \citep[see][for more specific details]{Del13}. For the present work this issue supposes a slightly easier background contribution to deal with for the PS detection methods with respect to the real maps. However, as the simulations are the same for the two methodologies, the comparison will still be a fair and informative one (see the end of section \ref{sec:results}).

Although generally negligible, for completeness we also take into account the thermal Sunyaev-Zel'dovich effect produced by galaxy clusters. Similarly to the Galactic emission, we use again the PLA simulations that are available for all \textit{Planck} channels. The simulations were produced following the \citet{Del02} method with a random distribution on the sky \citep[see][for more specific details]{Del13}.

Finally, we add the instrumental noise to the simulations. The noise maps are produced by simulating white noise according to the Planck values: 0.55, 0.78, and 2.56 $\mu K_{CMB}$ deg, respectively \citep[noise rms computed after smoothing to $1^\circ$;][]{PLA18_I}.

In this work we study the performance of two detection methods, PoSeIDoN and the MHW2, especially focusing on their dependence with increasing intensity of Galactic emission by applying two different homogeneous galactic cuts (at 10$^\circ$ and 30$^\circ$ Galactic latitudes). Moreover, a similar intensity increase also arises with higher frequencies due to Galactic emission spectral behaviour \citep{PEPXIX,PLA14_11, PLA16_10, PLA18_IV}.

Examples of random simulated patches are shown in the first two columns of Fig. \ref{Fig:patches} for 143, 217, and 353 GHz (top, middle, and bottom panels, respectively) at $b>30^\circ$ Galactic latitudes. The first column is the total input simulated map, including CMB, Galactic emission, CIB, PSs, and instrumental noise, whereas the second column is the input PS-only map.

\section{Methodology}
\label{sec:methodology}
\subsection{PoSeIDoN}
\label{sec:FCNN}
   \begin{figure*}[ht]
   \centering
   \includegraphics[width=17cm]{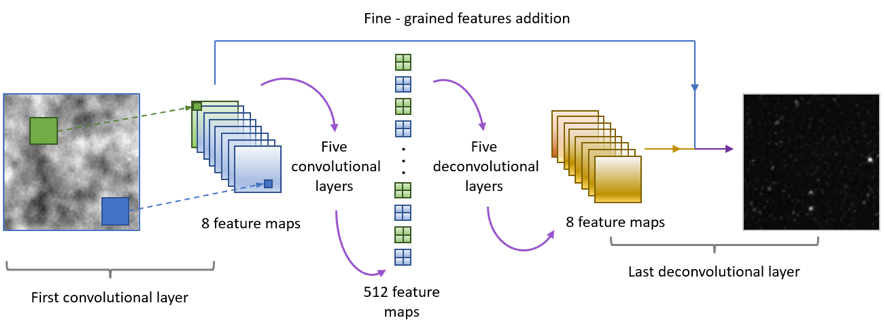}
      \caption{Details of the FCN used for PS detection in PoSeIDoN. The network has a block of eight convolutional layers, where the main characteristics are extracted, resulting in 512 feature maps, connected with a deconvolutional block of eight deconvolutional layers. Fine-grained features are added from each convolution to the corresponding deconvolution.}
         \label{Fig:FCN}
   \end{figure*}
Neurons, sorted in layers, are the basic computing elements of an artificial network model. Their responses are modelled by weights that represent the influence of the neuron response on the neurons of the subsequent layer. In particular, for some models (such as CNNs) the weights correspond with kernel values \citep{LEC15}. 
The kernels are tensor-shaped arrays that model the connections between neurons. These connections are then applied as discrete convolutions to all the inputs. Then, the feature maps obtained for each kernel become the inputs of the subsequent layers. The response is finally given after the process is completed along each computation unit.

In supervised learning, the implementation of the training procedure is performed via estimation of a loss function, usually a mean square error (MSE) function, computed over the data from a training set (i.e. the network responses to certain inputs compared with their corresponding labels). Back-propagation algorithms are then employed to correct weights and kernel values, and thus minimise the loss function with methods like the stochastic gradient descent (SGD) \citep{RUM88,CHA13}.

Fully convolutional networks allow us to perform dense predictions over the data used as input \citep{LON15, DAI16}. In this case the most relevant characteristics are first extracted using a convolutional block where each convolutional layer allows the extraction of several feature maps from the image obtained using kernels, frequently modulated by an activation function and processed by a down-sampling in terms of pooling. In addition to the typical convolutional process, an FCN has a second block where deconvolutions are performed, allowing the recovery of a dense response, also by means of layers with the correspondent kernels. Moreover, during the deconvolution process information on the convolutional segment is included through the addition of fine-grained features in specific steps.

In this work we considered the FCN approach because the attempts with other neural network models, such as MLP and CNN, although promising for simplified cases, had a clear limitation: the variable number of sources that could be found on the treated or processed image.
The search for the model was done from scratch, selecting the FCN parameters and hyperparameters through a grid search. Through the network search, special care was taken to avoid under-fitting and over-fitting, as under-fitted networks tend to detect fewer sources, and over-fitted networks tend to detect more than the actual number of sources. The selected topology is detailed as follows (see Figure \ref{Fig:FCN}):

\begin{itemize}
    \item Convolutional block: The network has six convolutional layers, with 8, 2, 4, 2, 2, and 2 kernels. The corresponding kernel sizes are 9, 9, 7, 7, 5 and 3 values per side. The activation function is leaky ReLU \citep{NAI10} in all the layers. Strides are of pixels both horizontally and vertically, and padding has been added.
    
    \item Deconvolutional block: The feature maps obtained after the convolutions are connected to a block of six deconvolutional layers. These layers have 2, 2, 2, 4, 2, and 8 kernels. The corresponding kernel sizes are 3, 5, 7, 7, 9 and 9 values per side. The activation function is leaky ReLU in all the layers. Strides are of pixels both horizontally and vertically, and padding has been added. Moreover, feature maps resulting from the five last convolutions are added, as fine-grained features, to the results of the five first deconvolutions.
\end{itemize}

The training procedure is performed using an MSE loss function, with a training set of 50000 samples and a validation set of 5000 samples. The test sets for performance assessment also consist of 5000 samples.
By analysing the evolution of the training loss function we decided to stop the training procedure at epoch 50. In the previous ten epochs, no relevant improvement was observed, with an average (standard deviation) MSE loss function value of $2.49\times10^{-3}$ ($5\times10^{-5}$). At these epochs the validation MSE loss function shows a constant behaviour with an average (standard deviation) value of $2.4\times10^{-3}$ ($2\times10^{-4}$). These values confirm that the network reached a good knowledge of the problem without any hint of over-fitting. Taking into account the averaged constant behaviour of the validation MSE loss function, it is probably an indication that we have reached its minimum value. Therefore, by continuing the training, we would only have obtained negligible improvements in the training MSE loss function while increasing the risk for over-fitting.

We produce 50000 simulations at 217 GHz to train the network. For each simulation we randomly chose a position of the available sky with the selected cut in latitude ($10^\circ$ or $30^\circ$) for both the CMB and the galactic emission, so that in each patch the background sky is never the same. Moreover, the positions and fluxes for the input PSs are also different in each realisation.

At this stage, for each patch, two images are provided to PoSeIDoN: the total image (the simulated patch including all the components, the `Input Total' column in Fig. \ref{Fig:patches}) and the PS image (the image containing only the input PSs that should be detected, i.e. all the radio and IR simulated PSs; the `Input PS' column in Fig. \ref{Fig:patches}). 
We note that  for  training purpose, the sources' flux density in the simulated catalogue are amplified by a training factor of 10, before being added to the other components. The reason is simply to increase the density of possible bright PSs inside the patch without modifying the source number count shape (i.e. without altering the statistical properties of the PS sample), just their normalisation. This step is not necessary or important, but it helps to reduce the training time as many of the sky patches do not have bright enough PSs to be detected due to their number density.

It should be noted that PoSeIDoN is trained just at 217 GHz and for the $30^\circ$ galactic cut. This trained FCN is then applied to all the cases studied in this work (i.e. 143, 217, and 353 GHz with a $30^\circ$ Galactic cut and 217 GHz with a $10^\circ$ Galactic cut). Better results are expected, although probably modest ones, if PoSeIDoN can be trained at each case individually. In this respect, the detection of PSs in regions with intense Galactic emission, such as the Galactic plane, is probably the most interesting case and also the one that can be improved the most by a dedicated FCN training. However, this is beyond the  scope of the current work. 

On the other hand, in the validation process, the simulated source flux densities are the realistic ones (no additional training factor is applied) that also allow us to compare our results with the Planck catalogues. 
The validation simulation is built using realistic PS flux densities and realistic contaminants simulated in the same way as for the training ones (although the sky positions are always randomly chosen). 
Each validation patch is then provided to the trained network that returns an output map of recovered PSs\footnote{There is no background or instrumental noise residual in the FCN output. Therefore, there is no detection threshold or uncertainty associated with detected sources.}. An example of the PoSeIDoN output patch at the studied frequencies is shown in Fig. \ref{Fig:patches}, last column. This output is then compared with the input PS-only map for a performance analysis: estimation of the completeness, reliability, and flux density accuracy. 

\subsection{Mexican hat wavelet 2}
To assess the performance of  PoSeIDoN  we also compare it to the MHW2 filter. 
The Mexican hat wavelet family in the plane is derived by applying the Laplacian operator iteratively to the 2D Gaussian \citep{GN06}. Any member of the family can be written in Fourier space as 
\begin{equation}
    \psi_n(k)=\frac{k^{2n}e^{-k^2/2}}{2^nn!}
.\end{equation}
The first member of the family, $\psi_1$, is the traditional MHW. It is one of the first wavelets applied successfully to the detection of PS in flat CMB maps \citep{CAY00, VIE01}. The MHW2 is therefore the second member of the family, $\psi_2$, and it was demonstrated to be even more suited to the task than its predecessor \citep{GN06} or the theoretical optimal matched filter \citep{TEG98}. It was successfully applied to the WMAP data \citep{GN08, MAS09} and it became the standard filtering technique for the production of the PS catalogues for the \textit{Planck} mission \citep{ERCSC, PCCS, PCCS2}. The wavelet coefficients, $w_n(b, R),$ can be obtained for each member of the family as
\begin{equation}
    w_n(b, R)=\int{dk e^{-ik·b}f(k)\psi_n(kR)},
\end{equation}
with $b$ being the location and $R$ the wavelet scale. 

By definition, the PSs adopt the beam profile or point spread function, usually approximated by a Gaussian,
\begin{equation}
\tau(x)= \frac{1}{2\pi\sigma_b^2} e^{-(\frac{x}{2\sigma_b})^2},    
\end{equation}
where $\sigma_b$ is the instrumental Gaussian beam dispersion. Therefore, the intensity of each source can be written as
\begin{equation}
I(x)=I_0 e^{-(\frac{x}{2\sigma_b})^2}.
\end{equation}

\noindent Then, the scale R of the wavelet can be optimised by finding the maximum signal-to-noise ratio (S/N) of the sources in the filtered patch, i.e. maximising the amplification factor $\lambda_n=\frac{w_n/\sigma_{wn}}{I_0/\sigma}$, with $\sigma$ and $\sigma_{wn}$ the rms deviation of the background before and after filtering, respectively. The optimal scale is determined for each patch independently and it is always near  unity.

The wavelets can be used for blind source detection (no prior information on the sources' positions) and in non-blind mode, usually to get the estimated flux densities of PSs at known positions. In our case we apply the filter blindly to each total input validation simulation to produce the filtered image. Examples of the MHW2 output image for our frequencies are the patches shown in Fig. \ref{Fig:patches}, third column.

\subsection{Catalogue production and statistical comparison}
The PoSeIDoN and the MWH2 methods both provide just an output image, not a list of detections. In this section we describe the catalogue production process and the statistical quantities that we use for the performance comparison.

The catalogue extraction consists simply in searching peaks (i.e. local maxima) above a certain intensity threshold, separated by at least a given minimum distance. This distance is 1.5 times the instrumental Gaussian beam dispersion, $\sigma_b$, which can be different for each channel.

Taking into account that the MHW2 is the most used technique to detect PSs in these kinds of images, we use it as our reference for the comparison. In particular, we use the standard deviation, $\sigma_{MHW2}$, of the MHW2 output map to set up the thresholds for catalogue production. In the case of the MHW2 we set a $3\sigma_{MHW2}$ threshold to build the catalogue of detected PSs (positions, flux densities, and uncertainties). To reduce the number of spurious sources we also apply the MHW2 with the flux density threshold set to 4$\sigma_{MHW2}$, reducing the completeness, as shown in section \ref{sec:results}. This last threshold level is approximately the same as used for the \textit{Planck} official PS catalogues \citep{ERCSC, PCCS, PCCS2}.

The input catalogue is built from the input PS-only image. Taking into account that this image does not have noise or any kind of background we use a lower PS threshold for the input catalogue. We choose one $\sigma_{MHW2}$ in order to be sure to have fainter PSs. In the PoSeIDoN case, the output map is an attempt to mimic the input PS image. It contains just PS candidates without any background or instrumental noise residuals. Therefore, we apply to the FCN output map the same procedure followed for the input catalogue (i.e. the 1$\sigma_{MHW2}$ threshold). 

A well-known issue of any filtering technique is the border effects. As can be seen in the MHW2 patches from Fig. \ref{Fig:patches} (third column), the filtering procedure produces artefacts near the patch border that can introduce spurious PS detections. 
In order to make a fair comparison with the MHW2 results, we exclude the detections within 5 pixels ($\sim 1$ FWHM) from the patches' borders on every side for all the cases (input, MHW2 and PoSeIDoN). 

To describe the performance of the two techniques we focus on three statistical quantities: completeness, number of spurious detections, and flux density estimation. These statistical quantities are commonly used to validate a detection technique or a produced catalogue \citep[see e.g.][]{LOP07, ERCSC, PCCS, PCCS2, PIPLIV, Hop15}.

Completeness is estimated by cross-matching the detected PSs against the input catalogue. It is a function of the intrinsic flux density, the detection threshold and sky location (not analysed in this work because the location dependency is common to both techniques by using the same simulations). Completeness provides information about the cumulative number of input sources that are missed at fainter flux densities: $C(>S_0)=\frac{N_{det}(>S_0)}{N_{input}(>S_0)}$, with $S_0$ the input flux density.

All the detection techniques misidentify background fluctuations as PSs at faint flux densities or around positions with strong Galactic emissions as the Galactic plane. Those incorrectly detected PSs, which are not in the input catalogue, are called spurious sources. Depending on the background characteristics and intensity, in many cases it is the number of spurious sources, and not the completeness, that puts a lower limit to the minimum flux density achievable with a given detection method. Therefore, the number of spurious sources is another important statistical quantity for the performance assessment of a detection technique.

Finally, the third statistical quantity is the flux density estimation. For these  PS detected input values, we can compare their flux densities. This comparison can provide useful information about potential flux density bias or can identify spurious sources that were detected by chance on the same positions of faint input PSs.


\section{Results}
\label{sec:results}

\begin{figure*}[h!]
   \centering
   \includegraphics[width=\columnwidth]{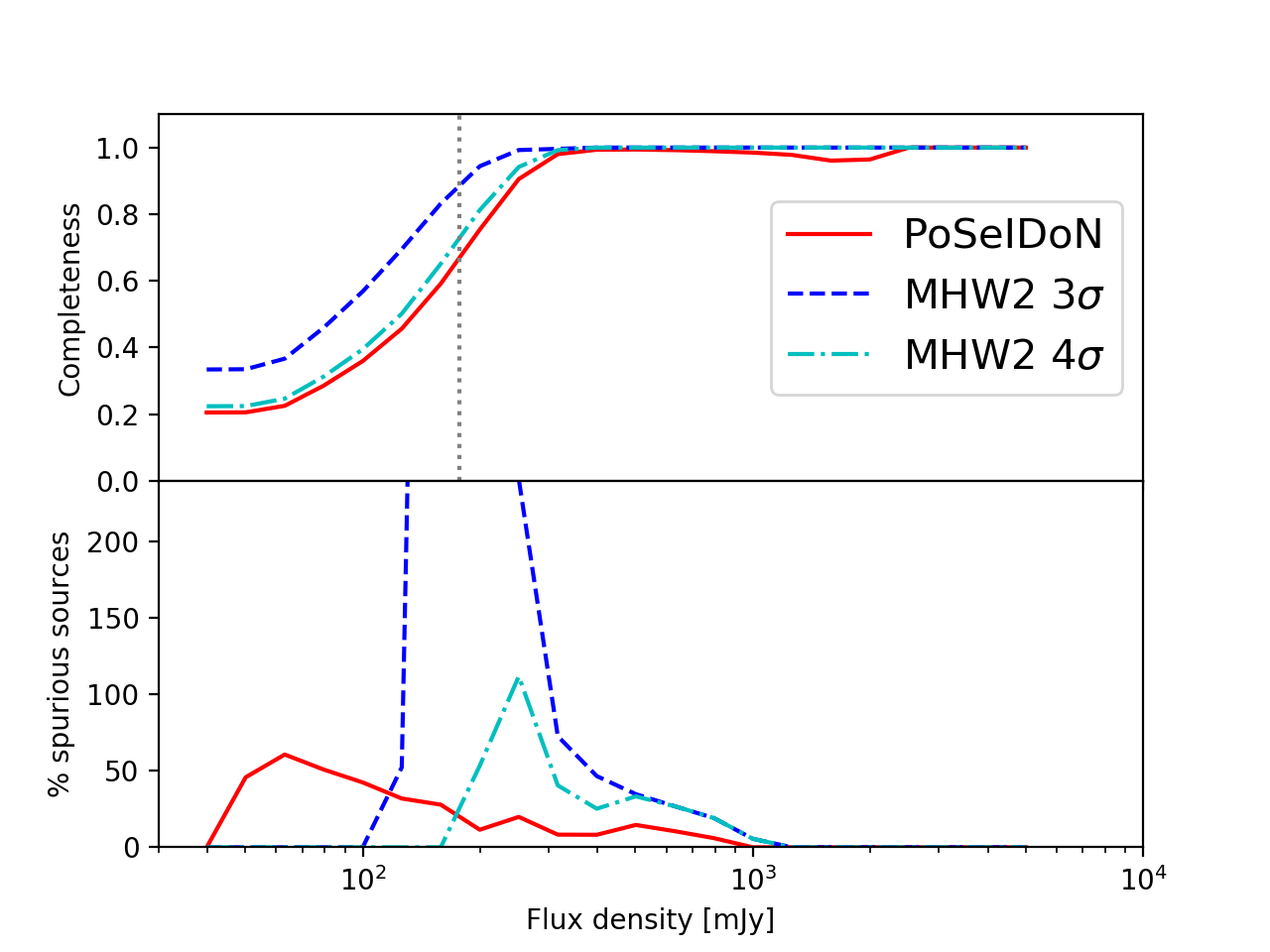} 
   \includegraphics[width=\columnwidth]{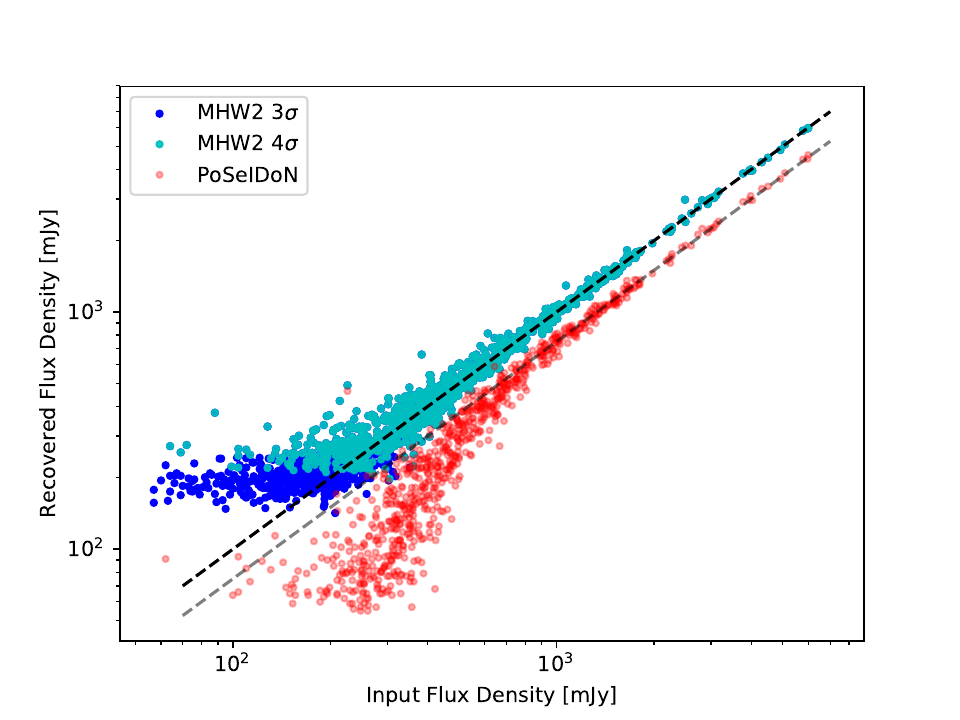}
   
   \includegraphics[width=\columnwidth]{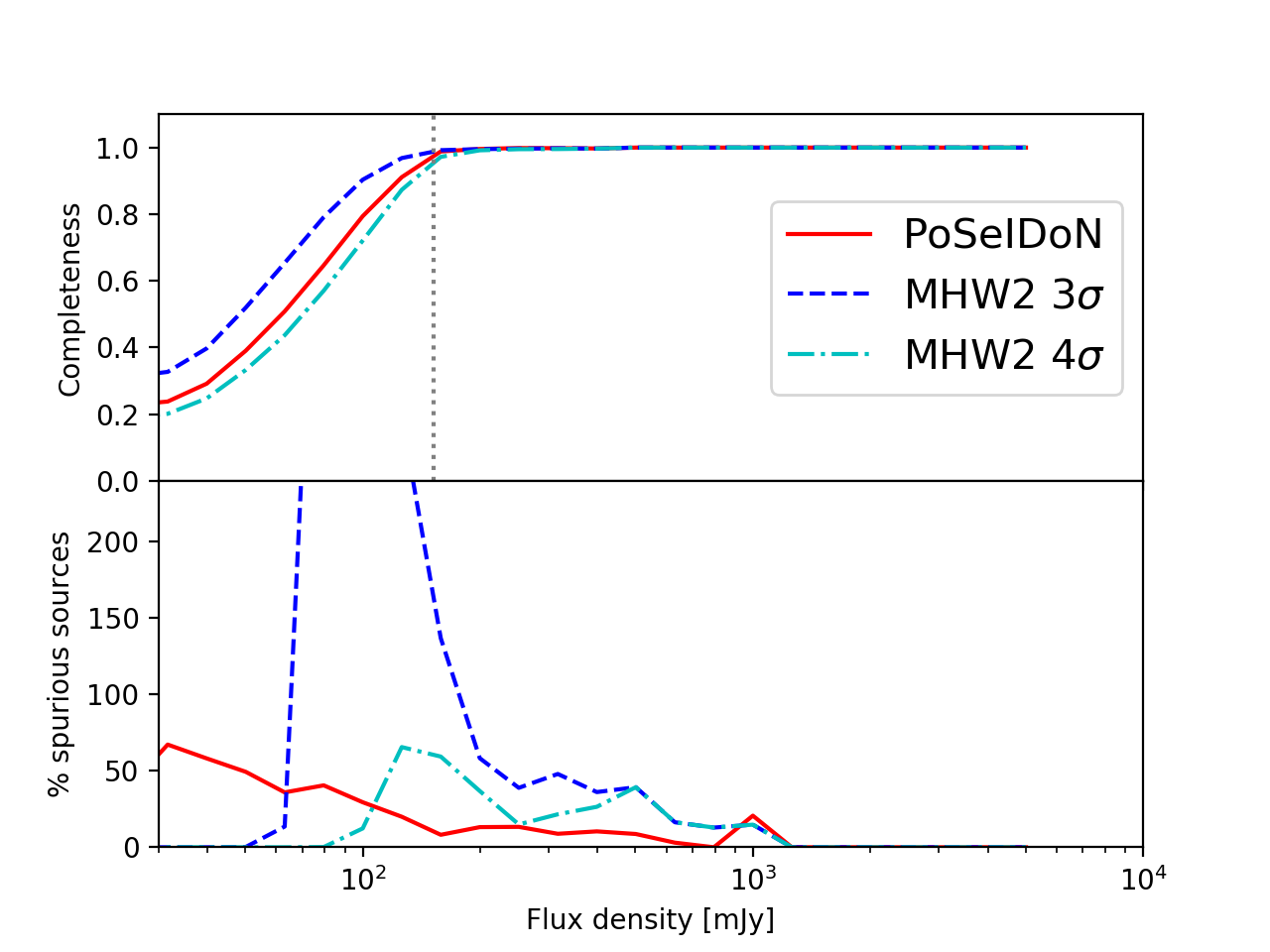} 
   \includegraphics[width=\columnwidth]{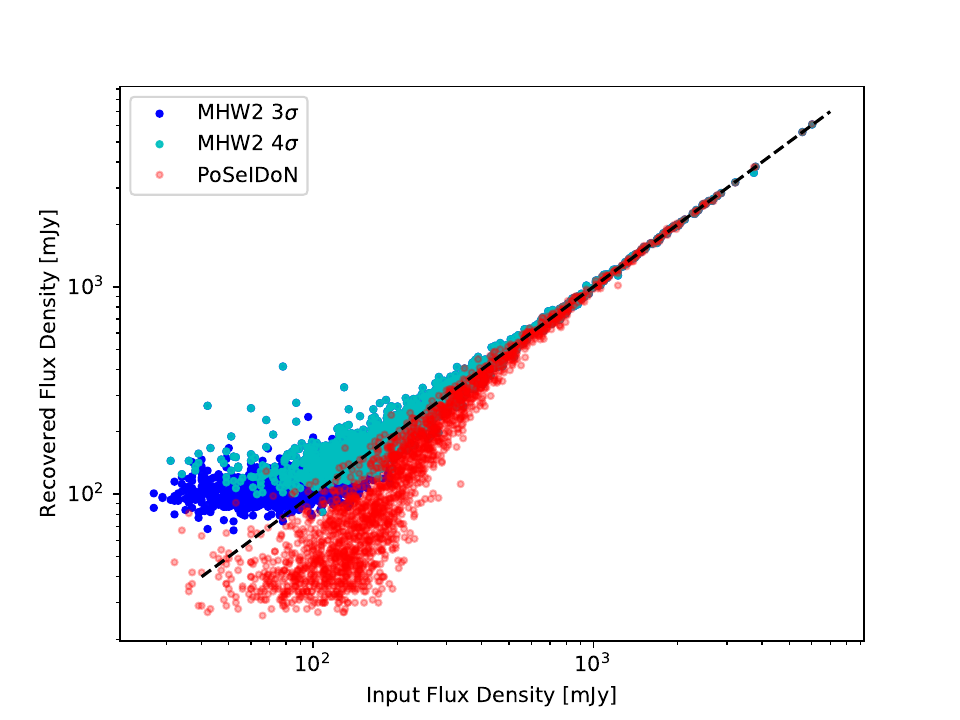}
   
   \includegraphics[width=\columnwidth]{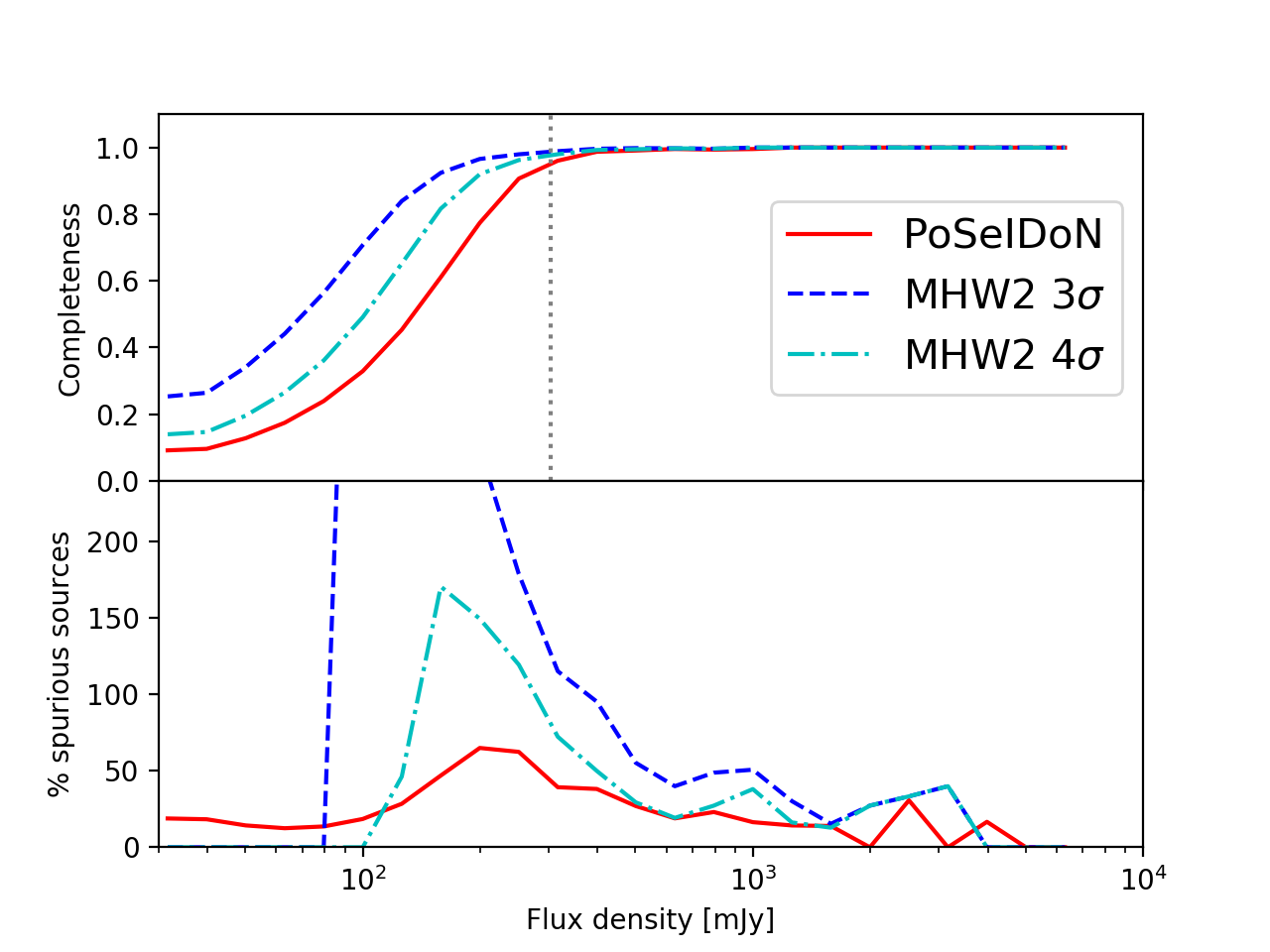} 
   \includegraphics[width=\columnwidth]{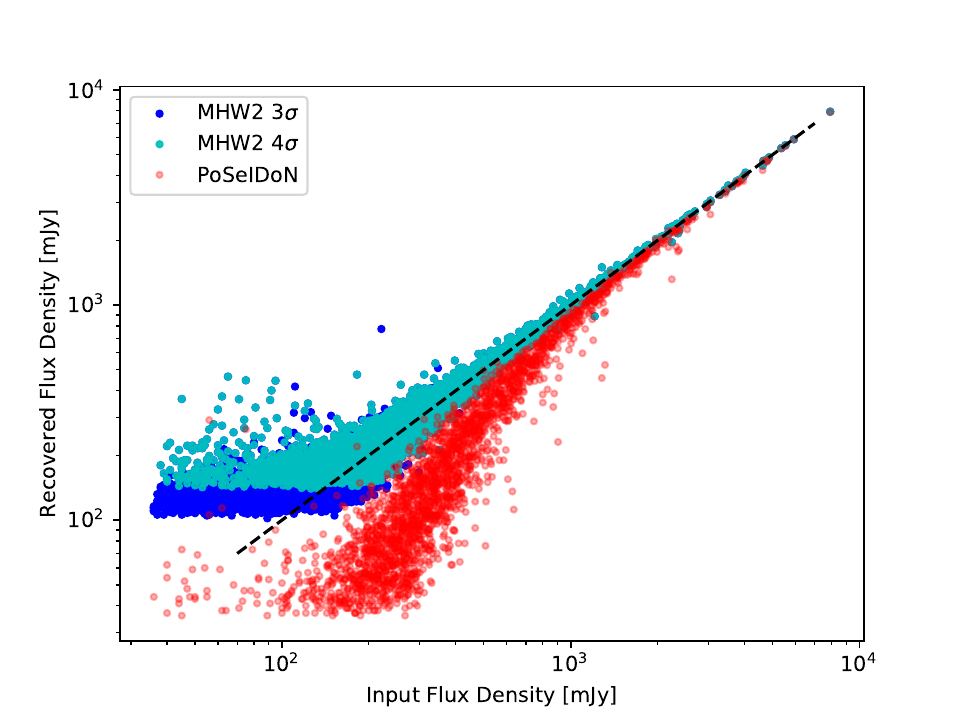}
   \caption{Validation results at 143 (top panels), 217 (middle panels), and 353 GHz (bottom panels) for a 30$^\circ$ galactic cut. Left column: Completeness (upper panel) and percentage of spurious sources (lower panel) estimated for PoSeIDoN (red solid line) and the MHW2 (blue dashed line for the 3$\sigma$ threshold and dot-dashed cyan for  4$\sigma$). The dotted grey vertical line is the 90\% completeness flux density limit for the PCCS2 \citep{PCCS2}. Right column: Flux density comparison between the input values and the recovered values by the three methods (same colors as in the left panels).}
\label{Fig:compl_rel}%
\end{figure*}

As mentioned above, Fig. \ref{Fig:patches} shows examples of output maps provided by the MHW2 and PoSeIDoN techniques when applied to the simulations. The third column corresponds to the MHW2, where   the typical granulated background after filtering can be seen. Moreover, these patches clearly show the border effects produced by the filtering approach. The fourth column  instead shows the PoSeIDoN output maps: no background fluctuations are present here, as the FCN only provides the best guess for input PSs. It should be noted that no border effects are present in the PoSeIDoN output images.

\subsection{Results at the training frequency: 217 GHz}
\subsubsection{Completeness and percentage of spurious sources}
As explained in section \ref{sec:FCNN}, we train the FCN only at 217 GHz for a Galactic mask of $|b|>30^\circ$ and we apply it to all the other studied cases. The performance of the two techniques at this frequency are compared in the middle panel of Fig. \ref{Fig:compl_rel} for the 30$^\circ$ galactic cut. The completeness (top sub-panel) and the percentage of spurious sources with respect to the input ones (bottom sub-panel) are shown on the left.

With the MHW2 we obtain the expected results: we have general agreement with previous applications of the MHW2 and in particular with the \textit{Planck} catalogues (where a $4\sigma$ threshold was used as a starting baseline; see the end of section \ref{sec:Completeness_others} for more details). Using a $3\sigma$ threshold (blue dashed line) the MHW2 provides good completeness results with a 90\% completeness level at 102 mJy. However, such an aggressive threshold implies a spurious PS detection problem already at $\sim500$ mJy with more than 20\% of the total detected sources being spurious ones. Using a more conservative and traditional $4\sigma$ threshold (cyan dot-dashed line), the spurious problem is highly reduced at least until $\sim300$ mJy. The price to be paid for this improvement is a reduction of the 90\% completeness level that increases to 137 mJy.

On the other hand, PoSeIDoN (red solid line) has a similar completeness performance with a 90\% completeness level of 126 mJy, between the results of the MHW2 with $3\sigma$ and $4\sigma$ thresholds. The clear advantage of the FCN is in the much lower number of spurious PSs; in this case it starts to be an issue only below $\sim100$ mJy (a flux density level below the 90\% completeness level). The spurious PS issue is strongly related with high intensity regions of the background (mainly the Galactic emission or a combination of CMB maxima and instrumental noise). Therefore, the fact that we have a lower number of spurious PSs shows a trend for PoSeIDoN in that it  more clearly distinguishes a PS from a background local maxima with respect to the general application of the MHW2.

\subsubsection{Photometry}
The recovered flux density is compared with the input value in the right column of Fig. \ref{Fig:compl_rel} (central panel for 217 GHz), providing additional information in the understanding of the finer PoSeIDoN performance when dealing with spurious PSs.

On the one hand, in the MHW2 case, most of the flux densities are correctly recovered within a 10\% relative error until $\sim200$ mJy. For the $3\sigma$ case (blue dots), at lower flux densities there are many sources whose recovered flux density is overestimated. Usually, these cases correspond to very faint input PSs that should not be detectable, which by chance are near the position of a positive CMB fluctuation or a region with strong Galactic emission \citep[it is the well-known `Eddington Bias';][]{EDD13}. Usually, these sources should be considered spurious, but in a real situation, without an input catalogue to be cross-matched, it is very difficult to identify them as they have flux densities always close to the detection limit for a wide range of input flux densities. As expected, this issue is less relevant with the more conservative $4\sigma$ threshold.

On the other hand, PoSeIDoN's behaviour on the recovery of the flux densities side is completely the opposite. The FCN recovers correctly the flux density of the  brightest sources, but tends to underestimate the flux density of fainter sources. Our interpretation for this specific behaviour is that the final flux density recovered by PoSeIDoN reacts as multiplied by a confidence factor, $S_{est}\propto p_{conf} S_0$. For bright input PSs or those in low background fluctuations areas, $p_{conf}\sim 1$ is used; on the contrary, for faint input PSs or those near high background fluctuations areas, $p_{conf}< 1$ is used.

This confidence factor has the advantage of  putting the most dubious detected PSs at fainter flux densities (see the steep increase of spurious sources below $\sim 150$ mJy), but it also means that the FCN recovered flux densities are not reliable. Although this is not ideal, it is not a limitation at all in the application of this novel technique; it is not unusual to first apply one technique for detection and then a second different one on the detected positions to estimate the flux density with better accuracy. One simple pipeline might consist in estimating the flux densities with the aperture flux, or with the MHW2 in non-blind mode, on the PS positions provided by PoSeIDoN. Another interesting possibility is to train a second neural network to get an accurate flux density estimation in known PS positions.

Moreover, by comparing the photometry of PoSeIDoN with that from a second method, a rough estimation of the reliability of an individual source can be obtained. Sources with high reliability will provide similar recovered flux densities for both methods, but if they differ by a great amount it means that $p_{conf}<< 1$ and therefore the source is probably a spurious one (or highly dominated by a background positive fluctuation).

\subsection{Results at non-training frequencies or different background properties}
\begin{figure*}[ht]
   \centering
   \includegraphics[width=\columnwidth]{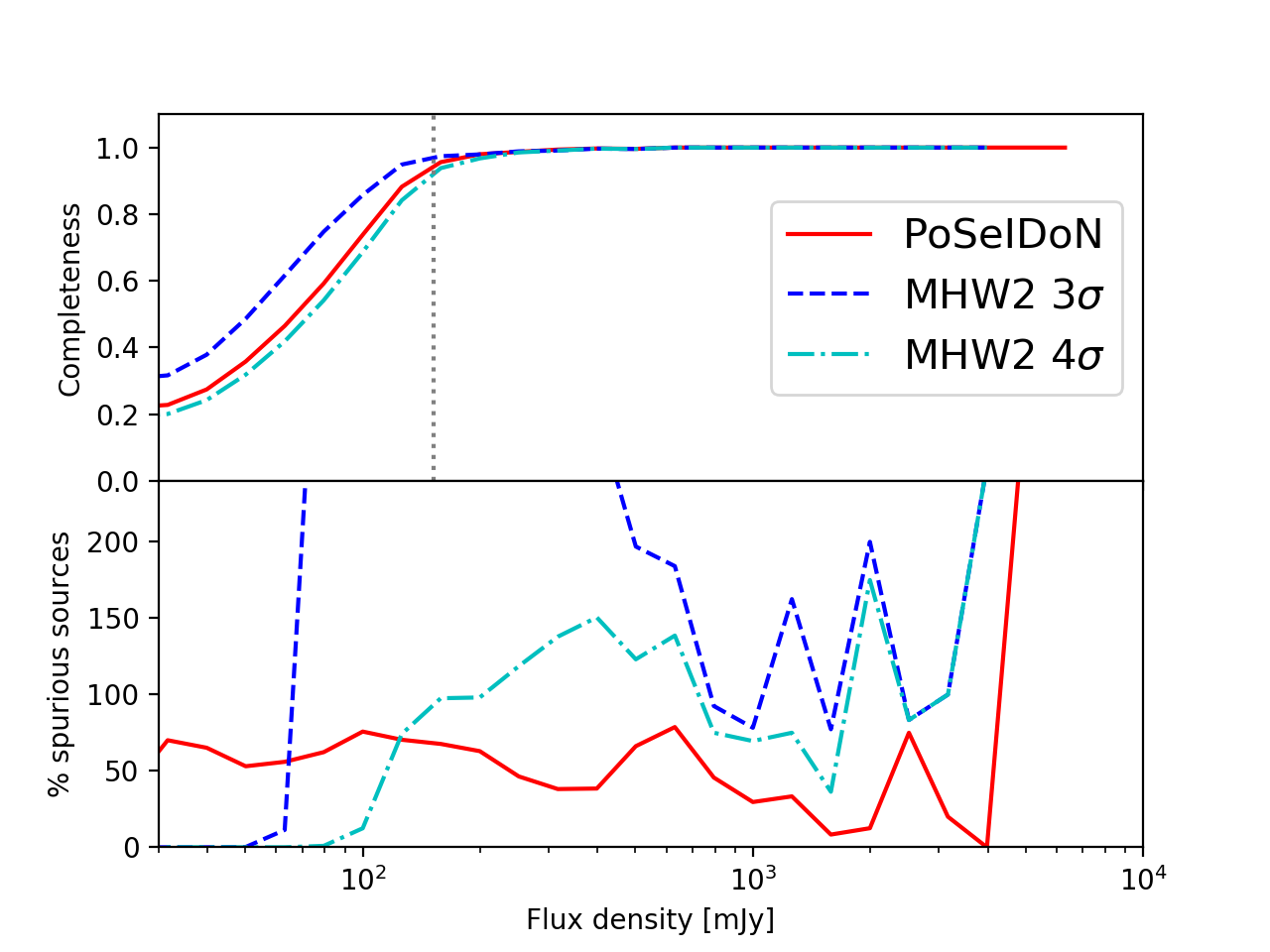}
   \includegraphics[width=\columnwidth]{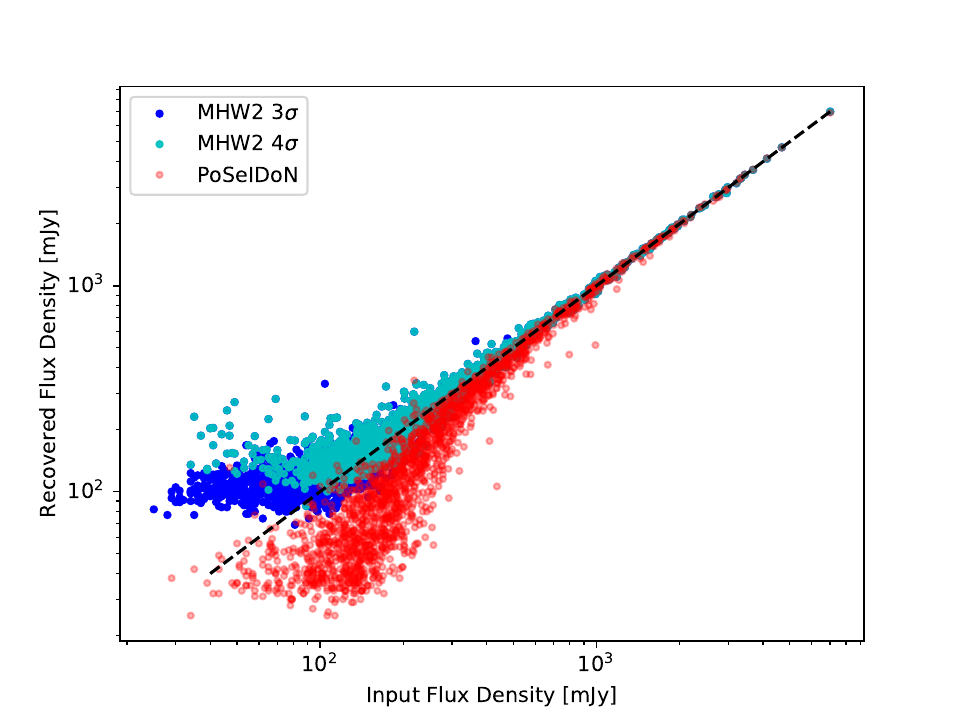}
      \caption{Validation results at 217 GHz for a 10$^\circ$ galactic cut. Left column: Completeness (top sub-panel) and percentage of spurious sources (bottom sub-panel) estimated for PoSeIDoN (red solid line) and the MHW2 (blue dashed line for the 3$\sigma$ threshold and dot-dashed cyan for  4$\sigma$). The dotted grey vertical line is the 90\% completeness flux density limit for the PCCS2 \citep{PCCS2}. Right column: Flux density comparison between the input values and the recovered values by the three methods (same colors as in the left panels).
      }
         \label{Fig:b10}
\end{figure*}
To test the robustness of PoSeIDoN we apply it to slightly different situations without additional training. On  the one hand, we apply it at 143 and 353 GHz (top and bottom panels of Figure \ref{Fig:compl_rel}) with the same galactic cut ($|b|>30^\circ$). The first channel has a lower Galactic emission but higher instrumental noise and bigger instrumental beam. The second channel has the same beam as 217 GHz but higher instrumental noise and the Galactic thermal emission is stronger. In addition, the FCN is applied again at 217 GHz, but allowing patches at lower Galactic latitudes, $|b|>10^\circ$, which implies stronger Galactic emission (see Fig. \ref{Fig:b10}).

\subsubsection{Completeness and percentage of spurious sources}
\label{sec:Completeness_others}
At 143 GHz (Fig. \ref{Fig:compl_rel}, top panel) the performance of PoSeIDoN with respect to the completeness is almost the same as the MHW2 with a $4\sigma$ threshold. The 90\% completeness level in this channel are 181, 235, and 253 mJy for the 3$\sigma$ MHWF, the 4$\sigma$ MHWF, and PoSeIDoN, respectively. In comparison with the 217 GHz case, the FCN performance has worsened with respect to that of  MHW2. The most probable reason is the change in the instrumental beam that will produce PSs that are slightly bigger than those used to train (and thus expected by) the FCN at 217 GHz. On the other hand, PoSeIDoN is still performing better regarding spurious PSs. While both MHW2 results have already more than 20\% spurious PSs at the 90\% completeness level, PoSeIDoN detects a lower number of spurious PSs well below $\sim$ 170 mJy.

At higher frequencies the Galactic emission is stronger and the spurious PS issue is worse. This is clearly shown in the spurious results of all the techniques at 353 GHz (Fig. \ref{Fig:compl_rel}, lower panel) that detect spurious sources above 1 Jy. PoSeIDoN  performs slightly better, although this issue is still present. At this frequency, a more conservative Galactic masking is needed or additional steps are required to decrease the number of spurious PSs \citep[see][]{PCCS, PCCS2}.
As for the completeness, the 90\% completeness levels are a bit higher with respect to the 217 GHz case, due to the higher background level: 153, 192, and 250 mJy for the 3$\sigma$ MHWF, the 4$\sigma$ MHWF, and PoSeIDoN, respectively. Again, as in the 143 GHz case, the PoSeIDoN completeness results are slightly worse than the 4$\sigma$ MHW2.

So, for all three methods (the 3$\sigma$ MHWF, the 4$\sigma$ MHWF, and PoSeIDoN) the source detection worsens at higher frequencies due to the increase in the foreground contribution to the total map.  PoSeIDoN is the overall preferable method, mainly due to the lower number of spurious sources.

To complete the analysis of the robustness of PoSeIDoN, we perform an additional test at 217 GHz. Without any additional training, we apply the FCN to a new set of validation simulations at 217 GHz but using a less aggressive Galactic mask of $|b|>10^\circ$ (see Fig. \ref{Fig:b10}). The completeness results remain more or less the same as in the 217 GHz case with $|b|>30^\circ$ for all the techniques. The 90\% completeness levels in this case are 110, 143, and 136 mJy for the 3$\sigma$ MHWF, the 4$\sigma$ MHWF, and PoSeIDoN, respectively.  However, the spurious PS numbers increase dramatically. Again, PoSeIDoN gives better results with this issue, and it shows a smoother increase in spurious PSs with decreasing flux density than in the 353 GHz case. This difference can be an indication that, as expected, the more closely the situation resembles the simulation training set, the better  the performance of the FCN is.

Therefore, by training PoSeIDoN in each particular situation the results can be slightly improved, although probably not by much in comparison with the MHW2. The detection of Galactic sources inside the complicated Galactic plane is most likely the most interesting case where the re-training would significantly improve the results. However, we demonstrate that even without specialised training, an FCN is able to compete with the MHW2 filtering scheme when applied to typical CMB experiment observed patches.

Finally, in the completeness panels of Figures \ref{Fig:compl_rel} and \ref{Fig:b10}, we  show (grey dotted line) the PCCS2 90\% completeness flux density limit \citep{PCCS2}: 177, 152, and 304 mJy at 143, 217, and 353 GHz, respectively. These values are in fair agreement with our findings, confirming that our simulations and detection procedures statistically resemble the Planck data. However, it should be noted that this information has been added to guide the reader and  is not meant as a direct comparison with our results. First of all, it must be taken into account that the PCCS2 is built to ensure at least 80\% reliability. Then, the different masking must also be considered. As for the percentage of masked sky, the PCCS2 excludes  15\%, 35.1\%, and 52.4\% for 143, 217, and 353 GHz \citep{PCCS2}, whereas our $30^\circ$ Galactic cut corresponds approximately to  50\% of the sky. Moreover, the PCCS2 masks are tailored to avoid the most contaminating Galactic areas and to maximise the sky coverage of the catalogue. We could have used a more effective masking, but for the purposes of  comparison between techniques  a simple galactic cut is sufficient. This point should  be taken into account when making a comparison with PCC2 numbers.

\subsubsection{Photometry}
As shown in the top and bottom panels of Figure \ref{Fig:compl_rel}, left column, the overall behaviour of all the methods' estimated photometry is approximately the same as in the 217 GHz case, although there are some differences.

At 143 GHz the MHW2 photometry seems to be less affected by the Eddington bias. This  probably occurs because of  the lower background level and higher instrumental noise that increase the detection level, but do not increase the number of high background areas. However, the PoSeIDoN photometry is highly affected even at the brightest PSs mainly due to the different instrumental beam with respect to the one used for training. The recovered flux densities above $\sim 400$ mJy are systematically 20\% lower than the input values (grey dashed line). For fainter sources, the recovered photometry shows the same $p_{conf}< 1$ behaviour as discussed in the 217 GHz case.

At 353 GHz the background level is higher  with respect to the 217 GHz value. For the MHW2 photometry the consequence is a much higher number of sources affected by the Eddington bias. As the instrumental beam is the same between 217 and 353 GHz, the PoSeIDoN photometry is in agreement with the input value for the brightest PSs. However, the stronger background makes $p_{conf}< 1$ at much brighter flux densities than before (i.e. due to the stronger background, PoSeIDoN is not completely certain about the detected PSs even above 800-1000 mJy).

However, this effect is more moderate in the $|b|>10^\circ$ case at 217 GHz. Again, this means that the performance of the FCN improves when the analysed image resembles the training images.

\subsection{ROC analysis}

  \begin{figure*}[ht]
   \centering
   \includegraphics[width=\textwidth]{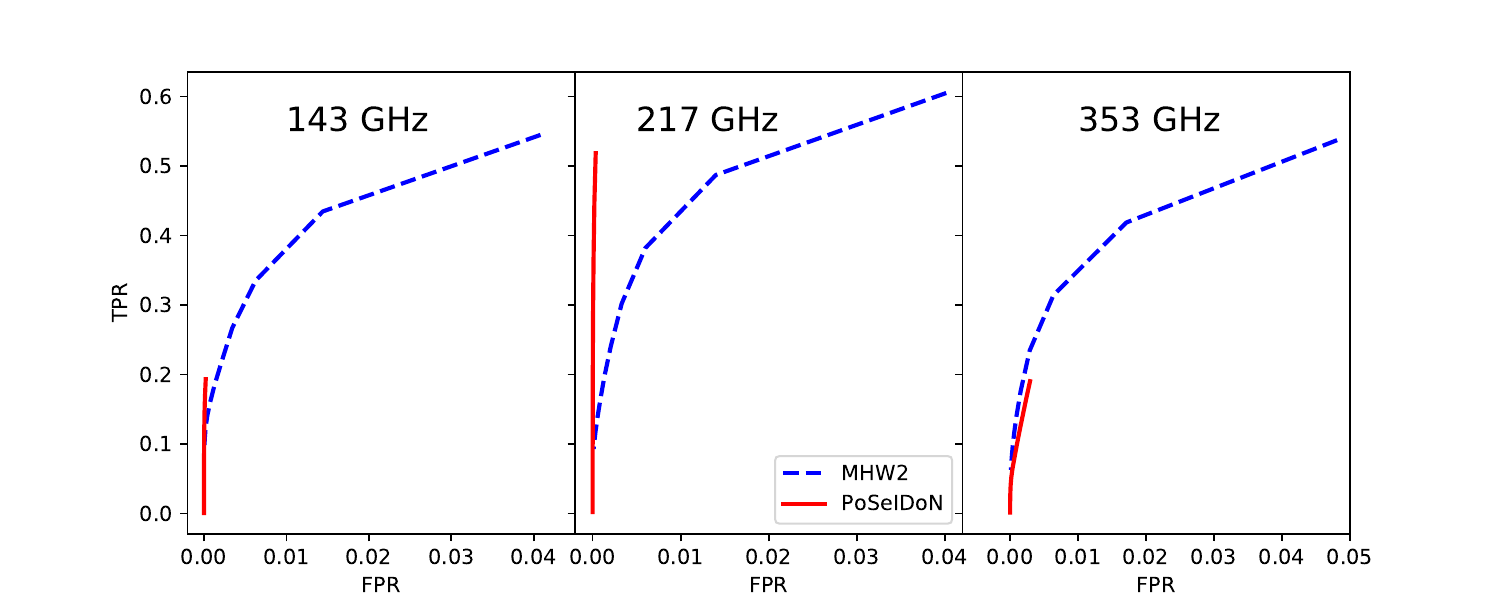}
   \caption{ROC plots for the studied methods at 143, 217, and 353 GHz (from left to right). In particular, PoSeIDoN is represented with the red solid line (using the flux density as a discriminating threshold), while the dashed blue line corresponds to the MHW2 starting at S/N=1.5.}
    \label{Fig:ROC}
   \end{figure*}

As we are comparing the performance of different PS detection methodologies, one useful quantity is the receiver operating characteristic  (ROC) curve \citep{KAY98}. ROC curves are widely used in  detection theory because they provide a direct and natural way to relate the costs and benefits of the decision making associated with the detection process in binary classifiers. The analyses performed during this work are done at the catalogue level. In order to perform the ROC analysis in this section we need to move to the pixel level where the definition of binary classifier applies completely. This means that for each set of images, we compare the pixels above a certain detection threshold (detected sources) with respect to the input image (point sources only).

The natural discriminating threshold to calculate the ROC curves should have been the S/N, but as explained before, the FCN provides an output image without any associated background or uncertainty. Therefore, we used the flux density in the FCN output images as the discriminating threshold to estimate the required quantities of the ROC analysis.

Thus, for each methodology we estimate the following quantities that are being used to create our ROC curves: the true positive rate (TPR) and the false positive rate (FPR). The TPR is defined as $TPR = \frac{TP}{TP+FN}$, where TP means true positive, or number of pixels of the output image above a certain threshold that correspond to pixels with sources in the input point source image, and FN means false negative, the number of pixels below the threshold that correspond to pixels with sources in the input point source image. On the other hand,  the FPR is related to the number of spurious sources or false positives, and is defined as  $FPR = \frac{FP}{FP+TN}$, where FP means false positive, or number of pixels of the output image above a certain threshold that does not correspond to pixels with sources in the input point source image, and TN means true negative, the number of pixels below the threshold that correspond to pixels without sources in the input point source image. All the ROC curves for the MHW2 have been calculated using S/N values from 1.5 (top right value) to 6.5 (bottom left value) with a 0.5 step. In the PoSeIDoN case, the flux density thresholds start at 50 mJy (top value) and  are increased with a logarithmic step of 0.1 until they reach 2 Jy (bottom value).

For any fixed FPR, the ROC curve gives the (normalised) number of detected pixels that coincide with pixels associated with input sources. ROC curves facilitate the comparison between two detectors: the higher curve in the plot is closer to the optimal performance than the lower curve. Figure \ref{Fig:ROC} shows the ROC curves for the two different methodologies at each of the three analysed frequencies (143 GHz, left panel; 217 GHz, central panel; and 353 GHz, right panel). All the methodologies have low TPR total values because the input catalogue has many faint sources not detectable by the methods. Furthermore,  the FPR total values are relatively low due to the sparsity of the point sources and therefore the high number of TN pixels.

Focusing firstly on the 217 GHz case, the typical behaviour for the MHW2 can be recognised. In order to detect more real PSs (i.e. increase the completeness or TPR), it needs to decrease the detection level that unavoidably increases the FPR (i.e. the spurious sources). 
However, as already discussed in the previous sections, the PoSeIDoN behaviour is very different. The maximum TPR is achieved almost without an important increase in the FPR value. The ROC curves simply confirm our previous conclusion about the good overall performance of PoSeIDoN thanks to the much better avoidance of spurious sources.

Similar conclusions can be derived at 143 GHz. However, the results of the two methods are more similar for low FPR values with respect to the 217 GHz case. This is due to the degradation of the PoSeIDoN performance when applied to a set of images different from the training ones. 

Finally, at 353 GHz the ROC curve for PoSeIDoN is slightly worse than that for  MHW2. Although the final number of spurious sources is much higher for the MHW2, it seems to be compensated by the steeper increase in the number of real detected PS. Considering again the observed degradation at other frequencies, at 353 GHz PoSeIDoN is penalised for being too conservative with slightly smaller completeness or TPR for a similar FPR with respect to the MHW2 (as discussed in section \ref{sec:Completeness_others}).

\subsection{Comparison with the PCCS2}

A full re-analysis of the PCCS2 is well beyond the scope of this paper. However, there is an interesting and useful comparison that can be easily performed. Taking into account the Galactic mask applied at 217 GHz during the production of the PCCS2, the simulations used to train PoSeIDoN should be realistic enough for a fair comparison.

First, we downloaded the actual \textit{Planck} total intensity map at 217 GHz\footnote{https://pla.esac.esa.int/\#home}. 
Then we projected a flat patch centred at each position of the 2135 PCCS2 listed sources at this frequency. The patch characteristics were the same as those used for the training of PoSeIDoN (128x128 pixels and a pixel size of 90 arcsec). The already trained PoSeIDoN (i.e. without any further adjustment or training from the simulations) was applied to each of the 2135 patches. For each of them,   only the detection at the centre was recorded, allowing a displacement of 5 pixels at maximum. As there is no uncertainty associated with the PoSeIDoN output images, we simply introduced a lower flux density limit of 10 mJy.  

\begin{figure}[ht]
   \centering
   \includegraphics[width=9cm]{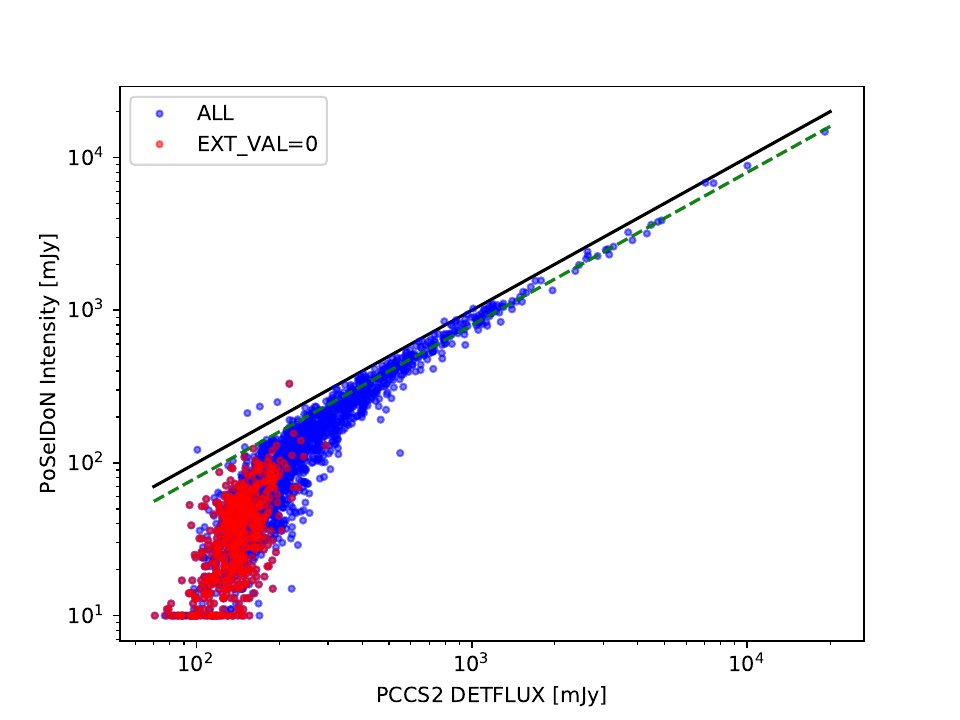}
      \caption{Comparison of the flux densities (DETFLUX) of the 2135 PCCS2 sources at 217 GHz and the intensities recovered by PoSeIDoN (blue circles). Those sources with the PCCS2 flag EXT\_VAL=0 are highlighted in red. The black solid line indicates the 1:1 relationship, while the green dashed line is the 10:8 (PCCS2 and PoSeIDoN, respectively).}
         \label{Fig:cnn_vs_PCCS2}
   \end{figure}

In Figure \ref{Fig:cnn_vs_PCCS2} the PCCS2 flux densities (DETFLUX) for the 2135 sources at 217 GHz are compared with the intensities recovered by PoSeIDoN (blue circles). As previously pointed out, PoSeIDoN intensities are the flux density of the sources but multiplied by a confidence factor. This peculiarity is also displayed in this comparison: the intensities from PoSeIDoN are $\sim20$\% fainter than the PCCS2 flux densities above 500 mJy (green dashed line). It is probably related to the differences between the real patches and the simulated ones \citep[e.g. the instrumental beam is not exactly a two-dimensional Gaussian, and it depends slightly on the position of the sky due to the scanning strategy][]{PEPVI}.

Below a flux density of 400 mJy, the intensities from PoSeIDoN become even fainter in comparison with the PCCS2 flux densities. This is an indication that the algorithm begins to be less confident on those detected sources (i.e. $p_{conf}< 1)$. Moreover, below $\sim200$ mJy, which corresponds to $\sim 160$ mJy of PoSeIDoN intensity, almost all the recovered intensities by PoSeIDoN are below the green dashed line. We cannot ensure that they are all spurious sources, but the fact that PoSeIDoN recovers much lower intensities than their flux densities is a strong warning that most of these sources are problematic.

In this respect the PCCS2 provides  flags that indicate the results from  external validations. An EXT\_VAL>0 means that the source was found in an external catalogue, in a previous version of the catalogue (ERCSC or PCCS), or  in another channel of the same PCCS2. The red circles shown in Figure \ref{Fig:cnn_vs_PCCS2} correspond to the PCCS2 sources at 217 GHz with EXT\_VAL=0 (i.e. sources only detected at 217 GHz in the PCCS2 version and therefore potentially spurious). Almost all of them have flux densities below 200 mJy and PoSeIDoN intensities are fainter than 100 mJy and clearly well below the green dashed line.

Based on our statistical results using the simulations during the validation of PoSeIDoN, we know that the percentage of spurious sources starts to increase for intensities below $\sim150$ mJy, being $\sim 25$ \% and $\sim 50$ \% for 100 mJy and 50 mJy, respectively (see left middle panel of Figure \ref{Fig:compl_rel}). This means that at least one-fourth of the flagged PCCS2 sources are probably spurious. Worse percentages are found if we consider the statistical results for the MHW2 4$\sigma$ with more than 50\% of spurious sources for flux densities below 200 mJy.

A detailed analysis for each individual source would be needed to confirm  whether they are spurious, but thanks to the better PoSeIDoN performance with respect to the spurious sources, we can confirm the PCCS2 warning that the flagged sources are likely to be spurious.

\section{Conclusions}
\label{sec:conclusions}

In this work we successfully applied PoSeIDoN to the detection of sources in a realistic situation; we included simulated PSs (both radio and IR) and CIB at the Planck frequencies of 143, 217, and 353 GHz; we added CMB, thermal SZ, and Galactic emission by randomly choosing the patches in the real \textit{Planck} CMB (provided by the \texttt{SEVEM} method) and the thermal SZ and Galactic simulated maps (provided by the PLA simulations); finally, we  added  instrumental noise according to the Planck characteristics.

The network was trained at 217 GHz with a Galactic cut of 30$^{\circ}$, using 50000 random simulations. Then it was applied to the validation simulations at 143, 217, and 353 GHz and Galactic cut of 30$^{\circ}$. At 217 GHz the network was also tested with a Galactic cut of 10$^{\circ}$. These results were then compared with those coming from the application of the MHW2 technique;  overall, \mbox{PoSeIDoN} is preferable because it provides more reliable results at lower flux densities (i.e. a lower number of spurious sources).

In should be noted that in the MHW2 case, in order to get rid of the many spurious sources detected at low fluxes, we need to increase the flux density detection limit from 3$\sigma$ to 4$\sigma$. On the contrary, the PoSeIDoN application is straightforward and  performs well; it achieves almost the maximum completeness or TPR with a very low number of spurious sources, as shown by the ROC analysis.

Another advantage of \mbox{PoSeIDoN} with respect to MHW2 is that it does not have border effects like any filtering approach. In the MHW2 analysis we need to remove those pixels near the patch border, subsequently missing those sources falling in that regions \citep[they can be recovered by selecting overlapping patches, as done in][]{ERCSC, PCCS, PCCS2}. \mbox{PoSeIDoN} is not affected by this problem, being able to detect sources placed near the limits of the patches.   

Moreover, we   highlight the good \mbox{PoSeIDoN} performance even at the frequencies where it was not trained. Therefore, even if the simulations used by future experiments are not accurate enough, an overall good PoSeIDoN performance is still expected.

As expected, the two methods worsen their performance with increasing frequencies, that is with the increase in the relative importance of the contaminants (mainly due to the Galactic thermal emission in our set of simulations). Moreover, they also get worse with a smaller galactic cut because the Galactic contamination is higher.

As a limit of PoSeIDoN, it must be said that the flux density estimation of the FCN method is not optimal, at least with respect to the MHW2; the network behaviour in flux density estimation is to give lower flux densities with respect to the true ones. We find a trend to assign lower flux densities to less reliable sources. So, our advice when building a catalogue (which is beyond the scope of this work) is to first blindly detect sources in a map with PoSeIDoN and then to estimate the flux density of the retrieved sources by non-blindly applying some flux density estimation methods (e.g. non-blind MHW2) in the obtained PoSeIDoN positions. The comparison between the recovered photometry of the two methods can also be used to get a rough estimation of the reliability of the individual PSs. A future development of the current work would be to train a second neural network to derive accurate flux density estimations in known PS positions.

During the analyses, we also indicated the flux density limit at 90\% completeness for the PCCS2, which is in fair agreement with our results. We applied PoSeIDoN (without any additional training) to the positions of the 2135 PCCS2 listed sources at 217 GHz. The results show that the intensities from PoSeIDoN are $\sim20$\% fainter than the PCCS2 flux densities above 500 mJy mainly due to the difference between the real instrumental beam and the Gaussian one used in the training simulations. The discrepancy in the photometry increases at fainter flux densities, which is an indication that PoSeIDoN is less confident in their PS nature. Most of these fainter sources are flagged as EXT\_VAL=0, which means that they are sources only detected at 217GHz in the PCCS2 version and therefore potentially spurious. Therefore, we can confirm the PCCS2 warning that the flagged sources are likely to be spurious.

As a final remark, we demonstrate that FCNs as PoSeIDoN over-performance can compete with and can be preferable to more traditional methods in detecting PSs embedded in noisy complex backgrounds in the microwave sky, as also shown by the ROC analysis. We demonstrate that, even just trained at one particular frequency and galactic latitude range, its performance is robust and competitive at different galactic latitudes or near frequencies. Obviously, the trained PoSeIDoN cannot be applied directly to other experiments with different characteristics (such as background level, beam sizes, or PS statistics). However, the PoSeIDoN FCN structure can be easily trained with specific simulations tailored for CMB experiments such as the Q-U-I JOint Tenerife Experiment \citep[QUIJOTE,][]{Rub12}, LiteBird, PICO or CMB-S4 ground-based experiments \citep[see e.g.][]{S4}. Even more, this kind of FCNs can also be trained to detect and characterise extended objects (shapes, intensity profiles) for missions like Euclid \citep{Lau11} or to decrease the deblending issue in overcrowded images for radio surveys, such as the Square Kilometre Array \citep[SKA;][]{DEW09}. 

\begin{acknowledgements}
LB, JGN and MMC acknowledge financial support from the PGC 2018 project PGC2018-101948-B-I00 (MICINN, FEDER). LB, JGN acknowledge financial support from PAPI-19-EMERG-11 (Universidad de Oviedo). JGN acknowledges financial support from the Spanish MINECO for the `Ramon y Cajal' fellowship (RYC-2013-13256). JDCJ, MLS, SLSG, JDS acknowledge financial support from the I+D 2017 project AYA2017-89121-P and support from the European Union's Horizon 2020 research and innovation programme under the H2020-INFRAIA-2018-2020 grant agreement No 210489629.\\
This research has made use of the python packages \texttt{ipython} \citep{ipython}, \texttt{matplotlib} \citep{matplotlib} and \texttt{Scipy} \citep{scipy}.
\end{acknowledgements}

%
%
\bibliographystyle{aa}
\bibliography{./SDNN}

\end{document}